\documentclass[superscriptaddress,twocolumn,amsmath,amssymb,aps,eqsecnum,prd,floatfix]{revtex4}
\begin{document}
\title{The Testbed for LISA Analysis Project}
\author{Lee Samuel Finn}
\email{LSFinn@PSU.Edu}
\altaffiliation{Department of Physics, Astronomy \& Astrophysics}
\affiliation{Center for Gravitational Wave Physics, 
The Pennsylvania State University, University Park, PA 16802, USA}
\author{Matthew J. Benacquista}
\email{benacquista@msubillings.edu}
\affiliation{Montana State University, Billings, MT 59101 USA}
\author{Shane L. Larson}
\email{shane@gravity.psu.edu}
\altaffiliation{Department of Physics}
\author{Louis J. Rubbo}
\email{rubbo@gravity.psu.edu}
\altaffiliation{Department of Physics}
\affiliation{Center for Gravitational Wave Physics, 
The Pennsylvania State University, University Park, PA 16802, USA}
\noaffiliation
\date{1 February 2006}

\begin{abstract}
The Testbed for LISA Analysis (TLA) Project  aims to 
facilitate the  development, validation and comparison of different 
methods for LISA science data analysis, by the broad LISA Science 
Community, to meet the special challenges that LISA poses. 
It includes 
a well-defined \emph{Simulated LISA Data Product} (SLDP), 
which provides a clean interface between the communities that have 
developed to model and to analyze the LISA science data stream; 
a web-based clearinghouse (at \texttt{<http://tla.gravity.psu.edu>})
providing SLDP software libraries, relevant 
software, papers and other documentation, and a repository for SLDP 
data sets; 
a set of mailing lists for communication between and among 
LISA simulators and LISA science analysts;
a problem tracking system for SLDP support; and 
a program of workshops to allow the burgeoning 
LISA science community to further refine the SLDP definition, 
define specific LISA science analysis challenges, and report
their results. 
This note describes the TLA Project, the resources it provides immediately, 
its future plans, and invites the participation of the broader community in
the furtherance of its goals. 

{\small\catcode`\$=12{}%
$Id: T06003.tex,v 1.4 2006/02/05 22:36:18 lsf Exp $
\catcode`\$=3}
\end{abstract}
\maketitle

\tableofcontents

\section{Introduction}

The Laser Interferometer Space Antenna (LISA) will be sensitive 
to gravitational waves arising from the coalescence of 
$10^4$--$10^8\,\textrm{M}_\odot$ black hole binaries in the centers 
of distant galaxies, the capture of stars, solar mass and intermediate 
mass black holes about these supermassive black holes, and the 
signature of hundreds of thousands of galactic binaries with periods 
ranging from hours to tens of seconds. 
The analysis of the LISA data stream to detect and learn from these 
sources poses unique and exciting statistical, computational, and 
algorithmic challenges. 
The Testbed for LISA Analysis (TLA) Project is intended to enable and 
facilitate the  development, validation and comparison of different 
methods for LISA science data analysis, by the broad LISA Science 
Community, to meet these challenges. 
This note describes the TLA Project and invites the participation of 
the broader community in its governance and its goals. 

\section{Goals, Objectives, Strategies}

The principal goal of the TLA is enable the burgeoning LISA science 
community to focus its energy on preparing to meet the challenges 
posed by LISA science analysis. 
A short-term objective, which advances us toward this goal, 
is to enable the science community to \emph{identify} the key LISA 
science data analysis challenges and \emph{provide proof-of-principle 
demonstrations} of different analysis techniques that meet these key 
challenges. 

To achieve its goals the TLA Project has identified the need for 
and provides several tools and an enabling infrastructure for collaboration. 
These include
\begin{itemize}
\item A web-based clearinghouse (\texttt{<http://tla.gravity.psu.edu>}) 
for sharing LISA data sets and 
related software, documentation and publications (and which also 
acts as a portal to ``all things LISA'') ; 
\item Several archived mailing lists for the discussion related to the use 
of these data sets and tools; 
\item A well-defined ``Simulated LISA Data Product'' (a container for simulated
LISA data that can be used to develop, validate and compare analysis 
methodologies), together with supported and maintained software libraries that 
can read and write the SLDP; 
\item A program of community open workshops and working meetings to further 
develop the SLDP, develop community-based analysis challenges, 
and facilitate their execution and the broad dissemination of their results; and
\item A governance structure that allows the science community to participate 
in the TLA at all levels, from setting the objectives and strategy that will move the 
TLA toward its goals, to contributing analysis or simulations tools, to helping to 
further develop the collaborative infrastructure. 
\end{itemize}
The following sections describe each of these components of the TLA. 

\section{The Simulated LISA Data Product and related software}

\subsection{Introduction}

Simulating LISA data and noise is an involved and complicated enterprise. 
There are different ways of representing the LISA science data (e.g., different 
TDI formulations), different approximations for constructing simulated 
LISA data (low-frequency, rigid adiabatic, etc.), several different LISA 
constellation ephemeris approximations (circular orbit, second-order eccentricity, 
etc.), and several different LISA simulators, each providing their output 
in a different format, etc. 

Nevertheless, the basic or schematic list of necessary and sufficient information 
required to carry-out LISA data analysis is quite short: 
\begin{itemize}
\item The constellation \emph{observables:} e.g., a specific set of TDI variables; 
\item The constellation \emph{response:} i.e., the relationship between the
observables and incident gravitational plane waves; 
\item The constellation \emph{ephemeris:} i.e., the position and orientation of the 
constellation as a function of time; 
\item A characterization of the constellation observable noise; and
\item Associated metadata (e.g., sampling rates for observables, ephemeris, 
time-varying response). 
\end{itemize}
If these are given then the particular choices made and details associated with 
the production of simulated LISA data sets are largely or entirely irrelevant to 
the data analyst.  To lower the barrier to community participation in the development, 
validation and comparison of different methods for LISA analysis it is thus crucial to 
separate the ``production'' of simulated LISA data from its ``use'' in science analysis. 

To make this separation and allow the processes of simulated data production 
and analysis to be carried out independently we have developed the 
Simulated LISA Data Product (SLDP). The SLDP is a well-defined container for 
the information, described above, that is necessary and sufficient for the development, 
validation and comparison of analysis methods. To facilitate the use of the SLDP
it is defined through an application programmer interface and the TLA Project 
provides and maintains implementations of that interface in the programming 
languages most commonly used for the production and analysis of simulated 
LISA data. The SLDP is the principal enabling technology of the TLA Project. 

In defining the interface between the simulator and the analyst as we have in the 
list above, we have made several implicit assumptions about the distinct responsibilities 
of the LISA Project and the LISA Science Analysis Community. In particular, we 
assume that when LISA is flying the LISA Project, and not the science analysis 
community, is responsible for initial processing of LISA data for the purpose of 
assuring the data integrity, characterizing the detector noise, identifying and removing 
instrumental artifacts, determination of the constellation response function, and 
preparing a structured data set that is ready for use in the scientific investigations 
that are the missions goal. While these are all important and challenging problems, 
they are distinct from the problem of analyzing the LISA data for gravitational wave 
sources, which is the thrust of the TLA Project.\footnote{As in any experimental 
enterprise, the lower the signal to noise the greater the 
degree of instrument familiarity required of any science team that presumes to 
carry-out science investigations with the instrument data. Nevertheless, as a 
practical matter there will always be a starting place for science investigations 
and this starting place will be a data product that is constructed, using the fullest 
awareness of the instrument's behavior, to be as clean of instrumental signatures 
or artifacts as the instrument team can make it. The SLDP is extensible and is
designed with the intent that future generations will have content that is increasing 
faithful to the complexity of a final LISA Data Product.}

The SLDP is intended to be an evolving standard. As LISA's design matures the
SLDP will also evolve to remain faithful to what we expect the data available for
science analysis will involve. Additionally, through the use of the SLDP we expect to 
learn more about how a LISA data product that facilitates the process of science analysis 
should be structured. 

\subsection{Implementing the SLDP}

Both the producers and users of SLDP data sets require a clean and well-defined 
software interface to the data sets: i.e., they should not need to be concerned with 
the underlying data set storage format. A suitable interface should be
\begin{itemize}
\item Well-defined;
\item Stable; 
\item Supported on all the major hardware platforms used for data set production 
  and analysis development;
\item Supported in all the major programming languages
  or environments used for data set production and analysis development; and 
\item Insulate both simulators and analysts from the underlying details of the storage 
  format.
\end{itemize} 
To lower the barrier to participation in the development of simulated LISA data and
the development of analysis methodologies the TLA Project provides, and is committed 
to continuing to provide and support, such an interface. 

While not of primary concern to the simulators or analysts, the underlying choice of 
data product storage format can have a significant impact on the success of this 
effort and the resources required to sustain it. The central objectives that governed 
the TLA's choice of data storage format were to minimize the barrier to participation 
presented by the storage format and minimize the effort required by the TLA Project in 
its development and maintenance. With these objectives in mind we imposed the 
following requirements on the data storage format:
\begin{itemize}
\item The storage format should be flexible enough to handle the typed and
  multi-dimensional array data objects that will represent LISA data; 
  
\item The storage format should be structured and self-documenting: 
  i.e., it should provide a table of contents of stored data objects, including 
  object descriptions and arbitrary user-defined object and data product
  metadata;
  
\item APIs (i.e., software libraries) for reading and writing files in this format
  should exist for all major programming languages, analysis development 
  environments, and hardware platforms;
  
\item The storage format definition, implementation and APIs should be
  well-documented, controlled and supported;
  
\item Storage format implementations and APIs that are controlled and
  supported by permanent organizations representing large and mature
  science communities are preferred over those that are not;
  
\item Storage format implementations and APIs that are free and freely
  distributable (i.e., no copyright and no export restrictions) are
  preferred over those that are not.
\end{itemize}

The version 5 Hierarchical Data Format (HDF5) (cf. 
\texttt{<http://hdf.ncsa.uiuc.edu/HDF5/>}) meets all the requirements posed 
above:
\begin{itemize}
\item HDF5 is a general purpose library and file format for storing
  scientific data. It is constructed about two primary objects:
  datasets, which are multi-dimensional arrays of data elements, and
  groups, which are hierarchically structured sets of datasets;
  
\item Datasets and groups are named and arbitrary metadata can be
  associated with each;
  
\item HDF5 APIs are available for C, C++, Fortran90, Mathematica, and
  Matlab, and the distribution is tested on Linux (ia64, x86\_64, and
  x86) and Mac OS X;
  
\item The HDF5 format and APIs are well-documented and designed for
  efficient storage and I/O in high performance, data intensive
  computing environments;
  
\item HDF5 was developed, and is maintained and supported, by the
  National Center for Supercomputer Applications (NCSA). It has a
  large and active user community that encompasses a large range of
  applications, including the packaging and distribution of NASA Earth
  Observing System datasets;
  
\item HDF5 is free and freely distributable in source form or as
  pre-compiled binary distributions for all major computing platforms.
\end{itemize}
SLDP data sets provided by the TLA will be in HDF5 format, 
include a TLA supported API, and implementations of the API for 
use in all major programming environments. 

\section{Collaborative Infrastructure: Web Clearinghouse, Mailing Lists 
and Workshop Program}

Communication and collaboration between and among simulators and 
analysts is critical for meeting the TLA Project goals. To support
this communication and collaboration the TLA Project includes a publicly
accessible web-based clearinghouse at \texttt{<http://tla.gravity.psu.edu>}
mailing lists to facilitate discussion amongst simulators and analysts; and 
a program of community-open workshops to enable the community to 
collaborate on the production of simulated LISA data, and 
development, validation and comparison of different analysis methodologies. 

The principal purpose of the TLA web site is to act as a clearinghouse for 
the distribution, exchange and support of SLDP data sets, the software 
required to read and write them, and any relevant documentation. In addition, 
The TLA web site will provide links to (and, on request, act as a repository for) 
any and all relevant software tools, documentation, technical notes, and science 
publication references related to the TLA's goals. Finally, the TLA site will also 
serve as a web-portal to ``all things LISA'', providing links to other LISA sites 
and sites relevant to LISA analysis goals and challenges. 


To share information of relevance to simulators and analysis developers and facilitate 
discussion between and amongst the members of these communities the TLA also 
supports several mailing lists. Initially four mailing lists are supported: 
\begin{itemize}
\item{} \texttt{<TLA-Simulate@Gravity.PSU.Edu>} is an unmoderated list where 
  questions and discussion related to the production of simulated LISA data can take place; 
\item{} \texttt{<TLA-Analyze@Gravity.PSU.Edu>} is an unmoderated list where questions
  and discussion related to the development, validation and comparison of simulated LISA 
  data can take place; 
\item{} \texttt{<TLA-Developer@Gravity.PSU.Edu>} is an unmoderated list where
  to support discussion amongst those taking part in the development of TLA Project
  support software; and
\item{} \texttt{<TLA-Announce@Gravity.PSU.Edu>} is a moderated list, deliberately 
  kept to a low volume, that provides announcements of broad interest to the broad 
  community interested in the TLA and its goals
\end{itemize}

Finally, the TLA Project will host a series of workshops and working meetings, intended 
to further the goal of enabling the burgeoning LISA science community to focus its 
energy on preparing to meet the challenges posed by LISA science analysis. Meetings
that are part of this program will focus on the simulation of LISA data; the development, 
validation and comparison of analysis methodologies; establishing contact and 
encouraging collaboration between the LISA science analysis community and the 
statistics, applied mathematics, and astrostatistics communities; and the development
and execution of analysis challenges, focused on demonstrating the capability of 
analysis technology in meeting LISA's science goals. 

\section{Getting Involved}

The goal of the TLA is to enable the burgeoning LISA science 
community to focus its energy on preparing to meet the challenges 
posed by LISA science analysis. Toward that end the TLA will do everything
possible to enable the participation of any and all members of the LISA 
science community in TLA activities. Interested members of the LISA Science 
Community can participate by 
\begin{itemize}
\item Simulating LISA data, with different levels of fidelity to the LISA design and 
expected gravitational wave sources, and 
providing data sets to be disseminated via the TLA web clearinghouse; 
\item Developing, validating and comparing data analysis methodologies by 
analyzing data sets and reporting their results; 
\item Contributing support software to enable the TLA Project goals; 
\item Providing their expertise to the broad community by taking part in the 
mailing list discussions; 
\item Taking part in workshops and working meetings devoted to LISA 
science analysis; and
\item Joining the \emph{TLA Development Team} and contributing to the development 
of TLA Project Infrastructure: API implementations in new languages; the design
and development of subsequent versions of the API, SLDP, and other collaborative
infrastructure; the administration of analysis drills and the evaluation of their results; etc. 
\end{itemize}
The TLA web site (\texttt{http://tla.gravity.psu.edu}) provides information on how to 
participate in all these ways. 

\section{Conclusions}

The goal of the Testbed for LISA Analysis (TLA) Project is to 
enable and facilitate the development, validation and comparison of 
different methods for LISA science data analysis. Its target audience is 
the broad LISA Science Community in general, and the communities of 
LISA science data modelers and LISA science data analysts in particular. 
To ease the exchange of information between these two communities, 
the TLA includes a well-defined Simulated LISA Data Product (SLDP); a 
web-based clearinghouse (<http://tla.gravity.psu.edu>) to provide SLDP 
data sets, software libraries, and other relevant software, papers and 
documentation;  a set of mailing lists to support communication among the 
sub-communities; and a problem tracking system for SLDP and related software 
support. In order to ensure that the TLA continues to grow and adapt to the 
needs of the broad scientific community, the TLA Project will support a 
program of workshops to allow the burgeoning LISA science community to 
further refine the SLDP definition, define specific LISA science challenges, 
and report their results.


\acknowledgments
The TLA project is supported by NASA awards NNG05GF71G and NNG04GD52G, 
The Pennsylvania State University Center for Space Research Programs, and The Center 
for Gravitational Wave Physics. The Center for Gravitational Wave Physics is supported by 
the National Science Foundation under cooperative agreement PHY 01-14375. 

\end{document}